\documentclass[12pt,english]{article}
\usepackage[english]{babel}
\usepackage[utf8]{inputenc}
\usepackage{natbib,authblk,bbm,bm}
\usepackage{amsmath,setspace,geometry,lineno,amsfonts}
\usepackage[vlined]{algorithm2e}
\SetKwInOut{Input}{Input}
\SetKwInOut{Output}{Output\,}
\SetKwInOut{Data}{Data}
\SetKwProg{Tree}{Tree}{}{EndTree}
\usepackage{graphicx}
\usepackage[colorinlistoftodos]{todonotes}
\usepackage{xcolor,colortbl}
\usepackage{fullpage}
\usepackage{mathtools}
\usepackage{framed}
\colorlet{shadecolor}{lightgray!20}
\usepackage{tikz}
\usepackage{bbm}
\usetikzlibrary{fit}
\usetikzlibrary{calc}
\usetikzlibrary{decorations.pathreplacing}
\usetikzlibrary{arrows.meta}
\tikzset{>={Latex[width=1.5mm,length=1.5mm]}}
\title{Gradient boosting in Markov-switching generalized additive models for location, scale and shape} 
\author{Timo Adam$^{1}\footnote{Corresponding author; email: \texttt{timo.adam@uni-bielefeld.de}.}$, Andreas Mayr$^2$ and Thomas Kneib$^3$\\ $^1$Bielefeld University, Germany\\ $^2$University of Bonn, Germany\\ $^3$University of Göttingen, Germany}

\begin{document}
\begin{spacing}{1.25}
\maketitle
\end{spacing}
\vspace{-9mm}
\begin{spacing}{1.5}

%%% abstract %%%
\begin{abstract}
We propose a novel class of flexible latent-state time series regression models which we call Markov-switching generalized additive models for location, scale and shape. In contrast to conventional Markov-switching regression models, the presented methodology allows us to model different state-dependent parameters of the response distribution --- not only the mean, but also variance, skewness and kurtosis parameters --- as potentially smooth functions of a given set of explanatory variables. In addition, the set of possible distributions that can be specified for the response is not limited to the exponential family but additionally includes, for instance, a variety of Box-Cox-transformed, zero-inflated and mixture distributions. We propose an estimation approach based on the EM algorithm, where we use the gradient boosting framework to prevent overfitting while simultaneously performing variable selection. The feasibility of the suggested approach is assessed in simulation experiments and illustrated in a real-data setting, where we model the conditional distribution of the daily average price of energy in Spain over time.
\end{abstract}

%%% keywords %%%

%%% section 1 -- introduction %%%
\section{Introduction}

In recent years, latent-state models --- particularly hidden Markov models (HMMs) --- have become increasingly popular tools for time series analyses. In many applications, the data at hand follow some pattern within some periods of time but reveal different stochastic properties during other periods \citep{zuc16}. Typical examples are economic time series, e.g.\ share returns, oil prices or bond yields, where the functional relationship between response and explanatory variables may differ in periods of high and low economic growth, inflation or unemployment \citep{ham89}. Since their introduction by \cite{gol73} nearly half a century ago, Markov-switching regression models, i.e.\ time series regression models where the functional relationship between response and explanatory variables is subject to state-switching controlled by an unobservable Markov chain, have emerged as the method of choice to account for the dynamic patterns described above.

While Markov-switching regression models are typically restricted to modeling the mean of the response (treating the remaining parameters as nuisance and constant across observations), it often appears that other parameters --- including variance, skewness and kurtosis parameters --- may depend on explanatory variables as well rather than being constant \citep{rig05}. A motivating example to have in mind is the daily average price of energy, which we present in detail in Section 5 for Spain as specific case study. When the energy market is in a calm state, which implies relatively low prices alongside a moderate volatility, then the oil price exhibits positive correlation with the mean of the conditional energy price distribution, but the variance is usually constant across observations. In contrast, when the energy market is nervous, which implies relatively high and volatile prices, then also the variance of energy prices is strongly affected by the oil price. This latter possible pattern cannot be addressed with existing Markov-switching regression models. As a consequence, price forecasts may severely under- or overestimate the associated uncertainty, by neglecting the strong heteroscedasticity in the process. This is problematic in scenarios where interest lies not only in the expected prices, but also quantiles, e.g.\ when the costs of forecast errors are asymmetric.

Since their introduction in the seminal work of \cite{rig05} a little more than a decade ago, generalized additive models for location, scale and shape (GAMLSS) have emerged as the standard framework for distributional regression models, where not only the mean, but also other parameters of the response distribution are modeled as potentially smooth functions of a given set of explanatory variables. Over the last decade, GAMLSS have been applied in a variety of fields, ranging from the analysis of insurance \citep{hel07} and long-term rainfall data \citep{vil10} over phenological research \citep{hud10} and energy studies \citep{vou11} to clinical applications, including long-term survival models \citep{dec10}, childhood obesity \citep{bey08} and measurement errors \citep{mayr17}.

GAMLSS are applied primarily to data where it is reasonable to assume that the given observations are independent of each other. This is rarely the case when the data have a time series structure. In fact, when the data are collected over time, as e.g.\ daily energy prices, then the functional relationship between response and explanatory variables may actually change over time. This results in serially correlated residuals due to an under- or overestimation of the true functional relationship. To exploit the flexibility of GAMLSS also within time series settings, we propose a novel class of flexible latent-state time series regression models which we call Markov-switching GAMLSS (MS-GAMLSS). In contrast to conventional Markov-switching regression models, the presented methodology allows to model different state-dependent parameters of the response distribution as potentially smooth functions of a given set of explanatory variables.

A practical challenge that emerges with the flexibility of MS-GAMLSS is the potentially high dimension of the set of possible model specifications. Each of the parameters of the response distribution varies across two or more states, and each of the associated predictors may involve several explanatory variables, the effect of which may even need to be estimated nonparametrically. Thus, a grid-search approach for model selection, e.g.\ based on information criteria, is usually practically infeasible. We therefore derive the MS-gamboostLSS algorithm for model fitting, which incorporates the gradient boosting framework into MS-GAMLSS. Gradient boosting emerged from the field of machine learning, but was later adapted to estimate statistical models \citep[c.f.][]{mayr14}. The basic idea is to iteratively apply simple regression functions (which are denoted as base-learners) for each potential explanatory variable one-by-one and select in every iteration only the best performing one. The final solution is then an ensemble of the selected base-learner fits including only the most important variables. The design of the algorithm thus leads to automated variable selection and is even feasible for high-dimensional data settings, where the number of variables exceeds the number of observations.

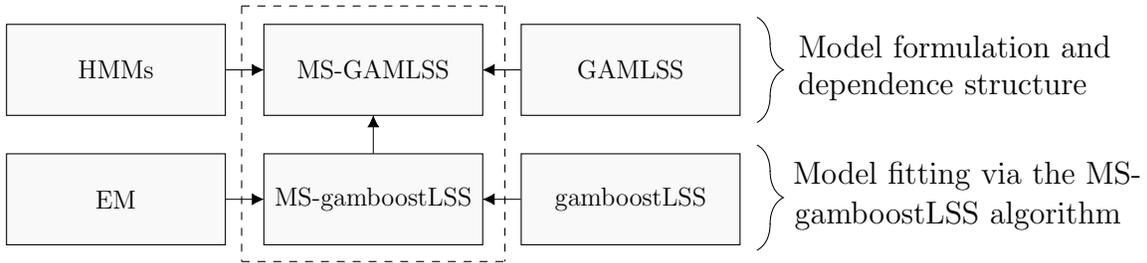
\begin{figure}[t!]
\centering
\begin{tikzpicture}[node distance = 2cm]
\tikzset{rect/.style = {rectangle, draw, minimum size = 42pt, minimum width={width("MS-gamboostLSS")+10pt}, scale = 0.82}}
\node [rect,fill=lightgray!10] (6) [] {MS-GAMLSS};
\node [rect,fill=lightgray!10] (9) [left = 5mm of 6] {HMMs};
\node [rect,fill=lightgray!10] (10) [below = 5mm of 9] {EM};
\node [rect,fill=lightgray!10] (20) [right = 5mm of 6] {GAMLSS};
\node [rect,fill=lightgray!10] (21) [below = 5mm of 20] {gamboostLSS};
\node [rect,fill=lightgray!10] (22) [below = 5mm of 6] {MS-gamboostLSS};
\draw[->, black, line width=0.2pt] (9) to (6);
\draw[->, black, line width=0.2pt] (20) to (6);
\draw[->, black, line width=0.2pt] (10) to (22);
\draw[->, black, line width=0.2pt] (21) to (22);
\draw[->, black, line width=0.2pt] (22) to (6);
\draw[dashed] (-1.75,0.85) -- (-1.75,-2.55);
\draw[dashed] (1.75,0.85) -- (1.75,-2.55);
\draw[dashed] (-1.75,0.85) -- (1.75,0.85);
\draw[dashed] (-1.75,-2.55) -- (1.75,-2.55);
\draw[decorate,decoration={brace,amplitude=10pt}] (5.1,0.7) -- (5.1,-0.7) node (6)[midway,below,]{};
\draw[decorate,decoration={brace,amplitude=10pt}] (5.1,-1.0) -- (5.1,-2.4) node (6)[midway,below,]{};
\node [text=black,] (1) [right = 27mm of 20] {};
\node [text=black,] (17) [above = -1.25mm of 1] {Model formulation and};
\node [text=black,] (18) [below = -1mm of 17] {\hspace*{-4.5mm}  dependence structure};
\node [text=black,] (19) [below = 11mm of 17] {\hspace*{2mm} Model fitting via the MS-};
\node [text=black,] (20) [below = -1.4mm of 19] {\hspace*{-0.5mm} gamboostLSS algorithm};
\end{tikzpicture}
\caption{In Section 2, we introduce the components of MS-GAMLSS and discuss the underlying dependence assumptions, which considers features from both HMMs and GAMLSS. In Section 3, we derive the MS-gamboostLSS algorithm, which incorporates gradient boosting into MS-GAMLSS.}
\label{fig1}
\end{figure}

The paper is structured as follows: In Section \ref{sec2}, we introduce the components of MS-GAMLSS and discuss the underlying dependence assumptions. In Section \ref{sec3}, we derive the MS-gamboostLSS algorithm and give a brief overview of related topics, including model selection. The synergy of HMMs and GAMLSS, which lies at the core of this work, is illustrated in Figure \ref{fig1}. In Section \ref{sec4}, we assess the suggested approach in simulation experiments, where we consider both linear and nonlinear base-learners. In Section \ref{sec5}, we illustrate the proposed methodology in a real-data setting, where we model the conditional distribution of the daily average price of energy in Spain over time.

%%% section 2 -- model formulation and dependence structure %%%
\section{Model formulation and dependence structure}
\label{sec2}

In this section, we introduce the model formulation and dependence structure of MS-GAMLSS, which extends the closely related but less flexible and in fact nested class of Markov-switching generalized additive models (MS-GAMs, \citealp{lan17}).

% the state process
\subsection{The state process}

MS-GAMLSS comprise two stochastic processes, one of which is hidden and the other one is observed. The hidden process, $\{ S_t\}_{t=1, \dots, T}$, which is referred to as the state process, is modeled by a discrete-time, $N$-state Markov chain. Assuming the Markov chain to be time-homogeneous, we summarize the state transition probabilities, i.e.\ the probabilities of switching from state $i$ at time $t$ to state $j$ at time $t+1$, in the $N \times N$ transition probability matrix (t.p.m.) $\boldsymbol{\Gamma}$, with elements
\begin{equation}
\gamma_{ij} = \Pr \left(S_{t+1} = j | S_t = i\right),
\label{gam}
\end{equation}
$i,j = 1, \dots, N$. The initial state probabilities, i.e.\ the probabilities of the process being in the different states at time 1, are summarized in the row vector $\boldsymbol{\delta}$, with elements
\begin{equation}
\delta_i = \Pr \left(S_1 = i\right),
\label{del}
\end{equation}
$i = 1, \dots, N$. If the Markov chain is assumed to be stationary, which is adequate in many applications, then the initial distribution is the stationary distribution, i.e.\ the solution to the equation system $\boldsymbol{\delta} \boldsymbol{\Gamma} = \boldsymbol{\delta}$ subject to $\sum_{i=1}^N \delta_i = 1$ \citep{zuc16}. If the Markov chain is not assumed to be stationary, then the initial state probabilities are parameters which need to be estimated. The state process is completely specified by the initial state and the state transition probabilities.

Throughout this paper we consider first-order Markov chains, i.e.\ we assume that the state process satisfies the Markov property, $\Pr(S_{t+1} | S_1, \dots, S_t) = \Pr(S_{t+1} | S_t)$, $t=1,\dots,T-1$. This simplifying dependence assumption is heavily exploited in the likelihood calculations provided in Section \ref{sec3}. While certainly being a strong assumption, in practice it is often a good proxy for the actual dependence structure, and could in fact be relaxed to higher-order Markov chains if deemed necessary \citep{zuc16}.

% the state-dependent process
\subsection{The state-dependent process}

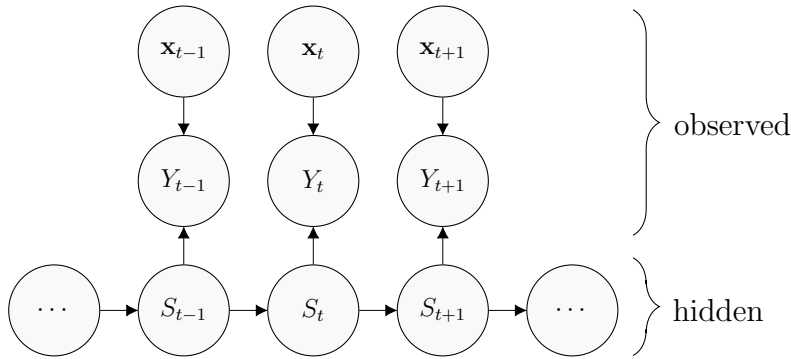
\begin{figure}[t!]
\centering
\begin{tikzpicture}[node distance = 2cm]
\tikzset{state/.style = {circle, draw, minimum size = 42pt, scale = 0.82}}
\node [state,fill=lightgray!10] (6) [] {$S_{t}$};
\node [state,fill=lightgray!10] (5) [left = 5mm of 6] {$S_{t-1}$};
\node [state,fill=lightgray!10] (7) [right = 5mm of 6] {$S_{t+1}$};
\node [state,fill=lightgray!10] (8) [above = 5mm of 5] {$Y_{t-1}$};
\node [state,fill=lightgray!10] (9) [above = 5mm of 6] {$Y_{t}$};
\node [state,fill=lightgray!10] (10) [above = 5mm of 7] {$Y_{t+1}$};
\node [state,fill=lightgray!10] (15) [left = 5mm of 5] {$\cdots$};
\node [state,fill=lightgray!10] (16) [right = 5mm of 7] {$\cdots$};
\node [state,fill=lightgray!10] (19) [above = 5mm of 8] {$\mathbf{x}_{t-1}$};
\node [state,fill=lightgray!10] (20) [above = 5mm of 9] {$\mathbf{x}_{t}$};
\node [state,fill=lightgray!10] (21) [above = 5mm of 10] {$\mathbf{x}_{t+1}$};
\node [text=black,] (17) [right = 23mm of 7] {hidden};
\node [text=black,] (18) [above = 19mm of 17] {\hspace*{2.5mm} observed};
\draw[->, black, line width=0.2pt] (15) to (5);
\draw[->, black, line width=0.2pt] (7) to (16);
\draw[->, black, line width=0.2pt] (5) to (6);
\draw[->, black, line width=0.2pt] (6) to (7);
\draw[->, black, line width=0.2pt] (5) to (8);
\draw[->, black, line width=0.2pt] (6) to (9);
\draw[->, black, line width=0.2pt] (7) to (10);
\draw[->, black, line width=0.2pt] (19) to (8);
\draw[->, black, line width=0.2pt] (20) to (9);
\draw[->, black, line width=0.2pt] (21) to (10);
\draw[decorate,decoration={brace,amplitude=10pt}] (4.25,0.7) -- (4.25,-0.6) node (6)[midway,below,]{};
\draw[decorate,decoration={brace,amplitude=10pt}] (4.25,4.0) -- (4.25,1.0) node (6)[midway,below,]{};
\end{tikzpicture}
\caption{Dependence structure in Markov-switching generalized additive models for location, scale and shape.}
\label{fig2}
\end{figure}

The observed process, $\{ Y_t\}_{t=1, \dots, T}$, which is referred to as the state-dependent process, can take either discrete or continuous values. We denote the conditional probability density function (p.d.f.)\ or, in the discrete case, probability mass function (p.m.f.), of $Y_t$, by
\begin{equation}
f_Y\left(y_t; \boldsymbol{\theta}_t^{(s_t)}\right) = f_Y \left(y_t; \theta_{1t}^{(s_t)}, \dots, \theta_{Kt}^{(s_t)}\right).
\label{dist1}
\end{equation}
Here $\boldsymbol{\theta}_t^{(s_t)} = (\theta_{1t}^{(s_t)}, \dots, \theta_{Kt}^{(s_t)})$ is the parameter vector associated with the distribution assumed for the response $Y_t$. It depends both on the current state, $s_t$, and on the explanatory variables at time $t$, $\mathbf{x}_t= (x_{1t}, \dots, x_{Pt})$, with $P$ denoting the number of variables included in the model. The first parameter of the response distribution, $\theta_{1t}^{(s_t)}$, often denotes the conditional mean of $Y_t$. Depending on the distributional family assumed, the other parameters may relate to the conditional variance, the conditional skewness and the conditional kurtosis, respectively, though other parameters are also possible. The set of possible distributions that can be specified for the response is not limited to the exponential family; in fact, any parametric distribution (including Box-Cox-transformed, zero-inflated and mixture distributions) can be considered. In principle, even more than four parameters could be considered, which however is of minor practical relevance and therefore omitted in the notation.

The variables $Y_1,\dots,Y_T$ are assumed to be conditionally independent of each other given the states and the explanatory variables, as illustrated in the graphical model depicted in Figure \ref{fig2}. As the parameters are possibly constrained (the conditional variance, for instance, typically needs to be strictly positive), we introduce a monotonic link function $g_{k}(\theta_{kt}^{(s_t)})$ for each parameter $\theta_{kt}^{(s_t)}$, $k=1,\dots,4$, which maps the latter onto some real-valued predictor function $\eta_{k}^{(s_t)}(\mathbf{x}_t)$, the choice of which is determined by the respective parameter constraints. For instance, the log-link function, $g_{k}(\theta_{kt}^{(s_t)}) = \log(\eta_{k}^{(s_t)}(\mathbf{x}_t))$, is typically chosen for the conditional variance, such that the inverse function, $\theta_{kt}^{(s_t)} = \exp (\eta_{k}^{(s_t)}(\mathbf{x}_t))$, is strictly positive. The form of the predictor function is determined by the specification of the base-learners, the discussion of which is subject of Section \ref{sec3.2}.

%%% section 3 -- model fitting %%%
\section{Model fitting}
\label{sec3}

In this section, we derive the MS-gamboostLSS algorithm to estimate the state transition probabilities (\ref{gam}), the initial state probabilities (\ref{del}) and the state-dependent parameters of the response distribution in (\ref{dist1}).

% the MS-gamboostLSS algorithm
\subsection{The MS-gamboostLSS algorithm}
\label{sec3.1}

The MS-gamboostLSS algorithm comprises an outer and an inner cycle, which combine two different model fitting procedures in a joint algorithm: The outer cycle is the EM algorithm (\citealp{dem77}; \citealp{bau70}; \citealp{wel03}), which is a popular method for iteratively maximizing the likelihood of a statistical model in the presence of missing data and has become one of the standard procedures for model fitting in HMMs. It is particularly useful in the context of MS-GAMLSS, as the hidden states can be regarded as missing data. The inner cycle is a weighted version of the gamboostLSS algorithm \citep{may12}, which is exploited to carry out one part of the EM algorithm, namely the estimation of the state-dependent parameters of the response distribution in (\ref{dist1}).

The missing data --- more precisely, functions of the missing data --- can be estimated, which is referred to as the expectation (E) step. Based on the obtained estimates, the complete-data log-likelihood (CDLL; i.e.\ the joint log-likelihood of the observations and and the states) is then maximized with respect to the state transition probabilities (1), the initial state probabilities (2) and the state-dependent parameters of the response distribution in (3), which is referred to as the maximization (M) step.

The complete-data log-likelihood: We represent the state sequence $\{ S_t \}_{t=1, \dots, T}$ (i.e., the missing data) by the binary random variables $u_i(t) = \mathbbm{1}_{S_t = i}$ and $v_{ij}(t) = \mathbbm{1}_{S_{t-1}=i, S_t=j}$ for $i,j = 1, \dots N$ and $t = 1, \dots, T$ (i.e., functions of the missing data). Assuming the $u_i(t)$'s and $v_{ij}(t)$'s to be observed, the CDLL can be written as
\begin{equation*}
\begin{split}
\text{CDLL} =& \log \left( \delta_{s_1} \prod_{t=2}^T \gamma_{s_{t-1} s_t} \prod_{t=1}^T f_Y \left( y_t; \boldsymbol{\theta}_t^{(s_t)} \right)\right) \\
=& \log \left(\delta_{s_1} \right) + \sum_{t=2}^T \log \left(\gamma_{s_{t-1} s_t} \right) + \sum_{t=1}^T \log \left( f_Y \left( y_t; \boldsymbol{\theta}_t^{(s_t)} \right) \right) \\
=& \underbrace{\sum_{i=1}^N u_i(1) \log \left(\delta_i\right) \vphantom{\sum_{j=1}^N}}_{\text{dependent on }\delta_i,~i=1,\dots,N} \hspace*{-3mm} + \underbrace{\sum_{i=1}^N \sum_{j=1}^N\sum_{t=2}^T v_{ij}(t) \log \left(\gamma_{ij}\right)}_{\text{dependent on }\gamma_{ij},~i,j=1,\dots,N} + \hspace*{0mm} \underbrace{\sum_{i=1}^N \sum_{t=1}^T u_i(t) \log \left( f_Y \left( y_t; \boldsymbol{\theta}_t^{(i)} \right) \right) \vphantom{\sum_{j=1}^N}}_{\text{dependent on } \eta_k^{(i)}(\mathbf{x}_t),~k=1,\dots,4}.
\end{split}
\label{cdll}
\end{equation*}
Note that the CDLL consists of three separate summands, each of which only depends on i) $\boldsymbol{\delta} = (\delta_i), i=1,\dots,N$, ii) $\boldsymbol{\Gamma} = (\gamma_{ij}), i,j=1,\dots,N$, and iii) $\boldsymbol{\theta}_t^{(i)} = (g_k^{-1}(\eta_k^{(i)}(\mathbf{x}_t))),i=1,\dots,N,k=1,\dots,4$, which considerably simplifies the maximization in the M step. Since the $u_i(t)$'s and $v_{ij}(t)$'s are not observable, we first need to replace them by their conditional expectations, $\hat u_i(t)$ and $\hat v_{ij}(t)$, respectively.

In order to compute these conditional expectations, we require the forward and backward probabilities: The forward probabilities, $\alpha_t(i) = f(y_1, \dots, y_t,$ $S_t = i | \mathbf{x}_1, \dots, \mathbf{x}_t)$, are summarized in the row vectors $\boldsymbol{\alpha}_t = (\alpha_t(1), \dots,$ $\alpha_t(N))$, which are calculated via the forward algorithm by applying the recursion
\begin{equation}
\begin{split}
\boldsymbol{\alpha}_1 &= \boldsymbol{\delta} \mathbf{P} \left(y_1\right)\\
\boldsymbol{\alpha}_t &= \boldsymbol{\alpha}_{t-1} \boldsymbol{\Gamma} \mathbf{P}\left(y_t\right),
\end{split}
\end{equation}
$t = 2, \dots, T$, where $\mathbf{P}(y_t) = \text{diag}(f_Y( y_t; \boldsymbol{\theta}_t^{(1)}), \dots, f_Y( y_t; \boldsymbol{\theta}_t^{(N)}))$. The backward probabilities, $\beta_t(j)$ $= f(y_{t+1}, \dots, y_T|S_t = j, \mathbf{x}_{t+1}, \dots, \mathbf{x}_T)$, are summarized in the row vectors $\boldsymbol{\beta}_t = (\beta(1), \dots, \beta(N))$, which are evaluated via the backward algorithm by applying the recursion
\begin{equation}
\begin{split}
\boldsymbol{\beta}_T &= \boldsymbol{1}\\
\boldsymbol{\beta}_t^\top &= \boldsymbol{\Gamma} \mathbf{P}\left(y_{t+1}\right)\boldsymbol{\beta}_{t+1}^\top,
\end{split}
\end{equation}
$t = T-1, \dots, 1$, with $\mathbf{P}(y_{t+1})$ as defined above. We let $\alpha_t^{[m]}(i)$ and $\beta_t^{[m]}(j)$ denote the forward and backward probabilities estimated in the $m^\text{th}$ iteration, which are computed using the predictors obtained in the $m-1$-th iteration (or offset values in the case of the first iteration).

The $m$-th E step involves the computation of the conditional expectations of the $u_i(t)$'s and $v_{ij}(t)$'s given the current parameter estimates, which leads to the following results:
\begin{enumerate}
\item[(1)] Since $\hat u_i(t) = \Pr(S_t = i|y_1, \dots, y_T, \mathbf{x}_1, \dots, \mathbf{x}_T) = f(y_1, \dots, y_t, S_t = i | \mathbf{x}_1, \dots, \mathbf{x}_T)$ $f(y_{t+1}, \dots, y_T|S_t = i, \mathbf{x}_1, \dots, \mathbf{x}_T) / f(y_1,\dots, y_T|\mathbf{x}_1, \dots, \mathbf{x}_T)$ and $f(y_1,\dots, y_T|\mathbf{x}_1, \dots,$ $\mathbf{x}_T) = \sum_{i=1}^N f(y_1,\dots, y_T, S_t=i |\mathbf{x}_1, \dots, \mathbf{x}_T)$, it follows immediately from the definition of the forward and backward probabilities that
\begin{equation}
\hat u_i^{[m]}(t) = \frac{\alpha_t^{[m]}(i) \beta_t^{[m]}(i)}{\sum_{i=1}^N \alpha_T^{[m]}(i)},
\label{u}
\end{equation}
$t = 1, \dots, T$, $i = 1, \dots, N$.
\item[(2)] Since $\hat v_{ij}(t) = \Pr(S_{t-1} = i, S_t = j|y_1, \dots, y_T, \mathbf{x}_1, \dots, \mathbf{x}_T) = f(y_1, \dots, y_{t-1}, S_{t-1} = i | \mathbf{x}_1, \dots, \mathbf{x}_T) \Pr(S_t = j | S_{t-1} = i) f(y_t, \dots, y_T | S_t = j, \mathbf{x}_1, \dots, \mathbf{x}_T) / f(y_1, \dots, y_T | \mathbf{x}_1, \dots,$ $\mathbf{x}_T)$, it follows immediately from the definition of the forward, backward and state transition probabilities that
\begin{equation}
\hat v_{ij}^{[m]}(t) = \frac{\alpha_{t-1}^{[m]}(i) \hat \gamma_{ij}^{[m-1]} f_Y \left( y_t; \hat{\boldsymbol{\theta}}_t^{(j){[m-1]}}\right) \beta_t^{[m]}(j)}{\sum_{j=1}^N \alpha_T^{[m]}(j)},
\label{v}
\end{equation}
$t = 1, \dots, T$, $i,j = 1, \dots, N$.
\end{enumerate}

The $m$-th M step involves the maximization of the CDLL with the $u_i(t)$'s and $v_{ij}(t)$'s replaced by their current conditional expectations with respect to the model parameters:
\begin{enumerate}
\item[(1)]{As only the first term in the CDLL depends on $\delta_{i}$, using a Lagrange multiplier to ensure $\sum_{i=1}^N \hat \delta_i^{[m]} = 1$ results in
\begin{equation}
\hat \delta_i^{[m]} = \frac{\hat u_i^{[m]}(1)}{\sum_{i=1}^N \hat u_i^{[m]}(1)} = \hat u_i^{[m]}(1),
\label{delta}
\end{equation}
$i = 1, \dots, N$.}
\item[(2)]{As only the second term in the CDLL depends on $\gamma_{ij}$, using a Lagrange multiplier to ensure $\sum_{j=1}^N \hat \gamma_{ij}^{[m]} = 1$, $i=1,\dots,N$, results in
\begin{equation}
\hat \gamma_{ij}^{[m]} = \frac{\sum_{t=2}^T \hat v_{ij}^{[m]}(t)}{\sum_{j=1}^N \sum_{t=2}^T \hat v_{ij}^{[m]}(t)},
\label{gamma}
\end{equation}
$i,j = 1, \dots, N$.}
\item[(3)] As only the third term in the CDLL depends on the state-dependent parameters of the response distribution in (3), the optimization problem effectively reduces to maximizing the weighted log-likelihood of a separate, conventional GAMLSS for each state, where the $t$-th observation is weighted by $\hat u_i^{[m]}(t)$. We can therefore exploit the gamboostLSS algorithm \citep{may12} to iteratively maximize this weighted log-likelihood. In particular, we consider the computationally more efficient non-cyclical variant of the gamboostLSS algorithm \citep{tom17}:

\begin{itemize}
\item \textbf{Initialize} the additive predictors $\hat\eta^{(i)[0]}_k(\mathbf{x}_t),i=1,\dots,N,k=1,\dots,4,t=1,\dots T$ with offset values. For each additive predictor, specify a set of base-learners $h_{k1}^{(i)}(x_{1t}), \dots, h_{kJ_k^{(i)}}^{(i)}(x_{J_k^{(i)}t})$ (e.g.\ simple linear models or penalized B-splines, i.e.\ P-splines; \citealp{eil96}), where $J_k^{(i)}$ denotes the cardinality of the set of base-learners specified for $\eta_k^{(i)}(\mathbf{x}_t)$.
\item For $i = 1$ to $N$:
\begin{itemize}
\item For $n=1$ to $n_\text{stop}^{(i)}$:
\begin{itemize}
\item For $k = 1$ to $4$:
\begin{itemize}
\item \textbf{Compute} the gradients of the CDLL with respect to $\eta_k^{(i)}(\mathbf{x}_t)$ (using the current estimates $\hat u_i^{[m]}(t)$ and $\hat{\boldsymbol{\theta}}_t^{(i)[n-1]} = (g_k^{-1}(\hat \eta_k^{(i)[n-1]}(\mathbf{x}_t)),k=1,\dots,4$),
\begin{equation*}
\nabla_{kt}^{(i)} = \frac{\partial \text{CDLL}}{\partial \eta_k^{(i)}\left(\mathbf{x}_t\right)} = \frac{\partial \sum_{t=1}^T \hat u_i^{[m]}(t) \log \left( f_Y \left(y_t; \hat{\boldsymbol{\theta}}_t^{(i)[n-1]} \right)\right)}{\partial \eta_k^{(i)}\left(\mathbf{x}_t\right)},
\end{equation*}
$t=1,\dots,T$, and \textbf{fit} each of the base-learners contained in the set of base-learners specified for $\eta_k^{(i)}(\mathbf{x}_{t})$ to these gradients.
\item \textbf{Select} the best-fitting base-learner $h_{kj^*}^{(i)}(x_{j^*t})$ by the residual sum of squares of the base-learner fit with respect to the gradients,
\begin{equation*}
j^* = \underset{j \, \in \, 1,\dots,J_k^{(i)}}{\operatorname{argmin}} \sum^T_{t=1} \left( \nabla_{kt}^{(i)}-\hat{h}_{kj}^{(i)}( x_{jt})\right)^2.
\end{equation*}
\end{itemize}
\item \textbf{Select}, among the base-learners selected the previous loop, the best-fitting base-learner $\hat h^{(i)}_{k^*j^*}(x_{j^*t})$ by the weighted log-likelihood,
\begin{equation*}
k^*=\underset{k \, \in \, 1,\dots,4}{\operatorname{argmax}} \sum^T_{t=1} \hat u_i^{[m]}(t) \log \left( f_Y \left( y_t; \hat{\boldsymbol{\theta}}_t^{(i)[n-1]} \right) \right),
\end{equation*}
where $\hat \theta_{kt}^{(i)[n-1]}$ is replaced by its potential update, $g_k^{-1}(\hat\eta^{(i)[n-1]}_k(\mathbf{x}_t) + \text{sl} \cdot\hat{h}_{kj^*}^{(i)}(x_{j^*t}))$, to \textbf{update} the corresponding predictor,
\begin{equation*}
\hat\eta^{(i)[n]}_{k^*}(\mathbf{x}_t) =\hat\eta^{(i)[n-1]}_{k^*}(\mathbf{x}_t) + \text{sl} \cdot\hat{h}_{k^*j^*}^{(i)}(x_{j^*t}),
\end{equation*}
where $0<\text{sl}<1$ is some small step length (typically, $\text{sl}=0.1$).
\end{itemize}
\item \textbf{Set} $\hat\eta^{(i)[n]}_k (\mathbf{x}_t) = \hat\eta^{(i)[n-1]}_k (\mathbf{x}_t)$ for all $k \neq k^*$.
\end{itemize}
\item \textbf{Use} the predictors obtained in the final iteration as estimates obtained in the $m$th M step, $\hat\eta^{(i)[m]}_k (\mathbf{x}_t) = \hat\eta^{(i)[n_\text{stop}^{(i)}]}_k(\mathbf{x}_t)$ for all $i,k$.
\end{itemize}
\end{enumerate}

The MS-gamboostLSS algorithm alternates between the E and the M step, each of which involves $n_\text{stop}^{(i)}$ boosting iterations for each state, $i$, until some convergence threshold, e.g.\ based on the difference between the CDLLs obtained in two consecutive iterations, is satisfied.

% specification of the base-learners
\subsection{Specification of base-learners}
\label{sec3.2}

The specification of base-learners, $h_{kj}^{(i)}(x_{jt})$, which are used to fit the gradient vectors, is crucial, as they define the type of predictor effect: If the base-learners have a linear form, then the resulting fit is also linear, whereas if nonlinear base-learners are chosen, then this fit may also be nonlinear. Generally, base-learners can be any kind of prediction functions --- in the classical machine learning context gradient boosting is most often applied with trees or stumps as base-learners \citep{gbm}. In the case of boosting algorithms for statistical modeling, it is, however, reasonable to select regression-type functions that can be combined to additive models \citep{mayr14}.

Due to their high flexibility, popular base-learners are P-splines \citep{eil96}. They are typically applied with fixed low degrees of freedom (strong penalization) which are not tuned for the different boosting iterations. However, as the same P-spline base-learner can be selected as best-performing base-learner and updated in several boosting iterations, the resulting solution can have arbitrarily large complexity (i.e.\ wiggliness). The complexity increases as the number of boosting iterations increases. More advanced base-learners are interaction terms (e.g.\ based on tensor product P-splines), random or spatial effects (e.g.\ based on Markov random fields). For an overview of available base-learners, see \cite{may12}.

% choice of the number of boosting iterations
\subsection{Choice of the number of boosting iterations}
\label{sec3.3}

The stopping iterations, $n_\text{stop}^{(i)}$, are the main tuning parameters for boosting algorithms. They control the variable selection properties of the algorithm and the smoothness of the estimated effects. They represent the classical trade-off between variance and bias in statistical modeling: Using more boosting iterations leads to larger and more complex models with smaller bias but larger variance, while stopping the algorithm earlier leads to sparser, less complex models with less variance but larger bias. Without early stopping, i.e.\ running the algorithm until convergence, the resulting fit converges to the maximum likelihood estimate \citep{may12} (if this estimate exists for the given model).

Choosing an optimal number of boosting iterations is typically achieved via $K$-fold cross validation. For some set $\boldsymbol{\Lambda} = \mathbf{n}_\text{stop}^{(1)} \times \dots \times \mathbf{n}_\text{stop}^{(N)} \subset \mathbb{N}^N$ we follow \cite{cel08} and proceed in the following way: First, we split the data into $K$ distinct partitions (typically, $K \geq 10$), estimate the model based on $K-1$ partitions and compute the out-of-sample log-likelihood for the remaining partition (which is straightforward using the forward algorithm from Section \ref{sec3.1}). This procedure is repeated $K$ times, i.e.\ until each partition has been out-of-sample once. The score of interest is the average out-of-sample log-likelihood over all partitions, where the number of boosting iterations corresponding to the highest score is chosen.

% selecting the number of states
\subsection{Selecting the number of states}

The choice of the number of states, $N$, is a rather difficult task --- while the vast majority of Markov-switching regression models appearing in the literature assume two states without any critical reasoning, there actually exists a variety of different methods for order selection in HMMs, which basically fall in two categories: On the one hand, a cross-validated likelihood approach can be used, as described in Section \ref{sec3.3}. On the other hand, information criteria such as Akaike's Information Criterion, the Bayesian Information Criterion \citep{zuc16} or the Integrated Completed Likelihood Criterion (\citealp{cel08}; \citealp{bie13}) can be considered, all of which result in a compromise between goodness of fit and model complexity.

One problem in practice, however, is that information criteria often tend to favor overly complex models. Real data typically exhibit more structure than can actually be captured by the model, which e.g.\ is the case if the true state-dependent distributions are too complex to be fully modeled by some (rather simple) parametric distribution or if certain temporal patterns are neglected in the model formulation. In the case of MS-GAMLSS, additional states may be able to capture this further structure. As a consequence, the goodness of fit increases, which may outweigh the higher model complexity. However, as models with too many states are usually difficult to interpret and are therefore often not desired, information criteria should be considered as a rough guidance rather than as a deterministic decision rule, which should be treated with some caution. For an in-depth discussion of pitfalls, practical challenges and pragmatic solutions regarding order selection in HMMs, see \cite{poh17}.

%%% section 4 -- simulation experiments %%%
\section{Simulation experiments}
\label{sec4}

To assess the performance of the suggested approach, we present two different simulation settings, where we consider linear (Section \ref{sec4.1}) and nonlinear (Section \ref{sec4.2}) relationships between the explanatory variables and the parameters of the response distribution.

% linear setting
\subsection{Linear setting}
\label{sec4.1}

For the linear setting, we use simple linear models as base-learners. In each of $100$ simulation runs, we simulated $500$ realizations from a 2-state Markov chain, $\{ S_t \}_{t=1,\dots, 500}$, with off-diagonal t.p.m.\ entries $\gamma_{ij} = 0.05$, $i,j = 1,2$, $i \neq j$, and initial state probabilities $\delta_i = 0.5$, $i=1,2$. Based on the simulated state sequence, we then draw $500$ observations from a negative binomial distribution with state-dependent p.m.f.
\begin{equation*}
f_Y\left(y_t; \theta_{1t}^{(s_t)},\theta_{2t}^{(s_t)}\right)=\frac{\Gamma \left(y_t+\theta_{2t}^{(s_t)}\right)}{\Gamma (y_t+1)\Gamma \left(\theta_{2t}^{(s_t)}\right)} \frac{\left( \frac{\theta_{1t}^{(s_t)}}{\theta_{2t}^{(s_t)}}\right)^{y_t}}{\left( \frac{\theta_{1t}^{(s_t)}}{\theta_{2t}^{(s_t)}+1} \right)^{\left(y_t+\theta_{2t}^{(s_t)}\right)}},
\end{equation*}
where
\begin{equation*}
\begin{split}
\log(\theta_{1t}^{(1)}) &= \eta_1^{(1)}(\mathbf{x}_t) = 2+ 2 x_{1t} + \sum_{j=2}^{100} 0 x_{jt} \\
\log(\theta_{1t}^{(2)}) &= \eta_1^{(2)}(\mathbf{x}_t) = 2 - 2x_{1t} + \sum_{j=2}^{100} 0 x_{jt}\\
\log( \theta_{2t}^{(1)} ) &= \eta_2^{(1)}(\mathbf{x}_t) = 2x_{1t} + \sum_{j=2}^{100} 0 x_{jt}\\
\log( \theta_{2t}^{(2)} ) &= \eta_2^{(2)}(\mathbf{x}_t) = -2 x_{1t} + \sum_{j=2}^{100} 0 x_{jt}
\end{split}
\end{equation*}
and $x_{jt} \sim \text{uniform}(-1,1)$, $j=1,\dots,100,t=1,\dots,500$. To assess the variable selection performance, we included $99$ noninformative explanatory variables in each predictor. The stopping iterations were chosen via $20$-fold cross validation over the grid $\boldsymbol{\Lambda}= \mathbf{n}_\text{stop}^{(1)} \times \mathbf{n}_\text{stop}^{(2)}$, $\mathbf{n}_\text{stop}^{(1)}=\mathbf{n}_\text{stop}^{(2)}= (100, 200, 400, 800)$, where the average chosen number of boosting iterations was $435$ (state 1) and $468$ (state 2).

\begin{figure}[t!]
\begin{minipage}{0.4875\textwidth}
\includegraphics[width=\textwidth]{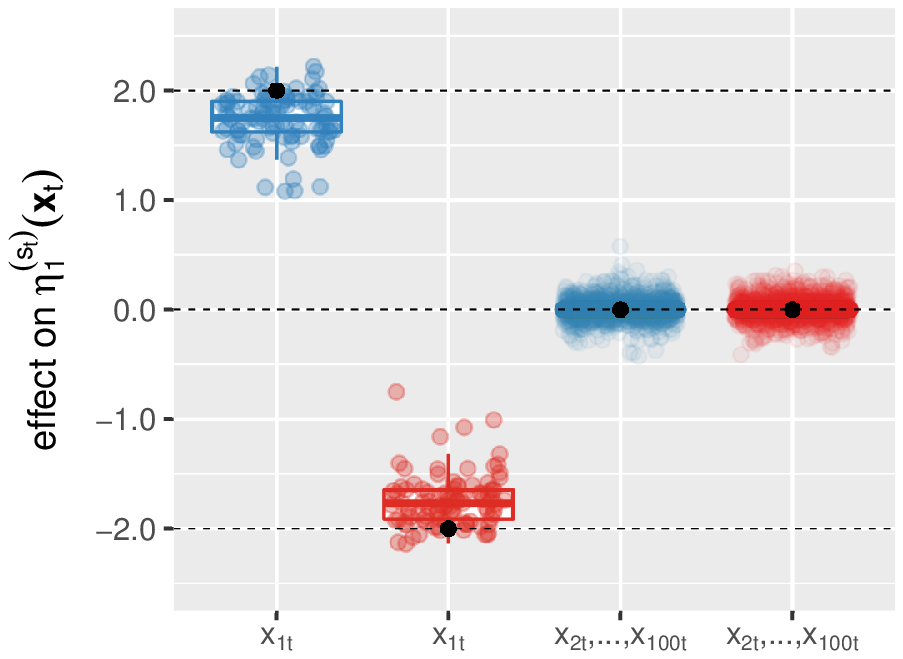}
\caption{Estimated state-dependent coefficients for $\theta_{1t}^{(s_t)}$ for state 1 (blue) and 2 (red) obtained in $100$ simulation runs. The true parameters (i.e.\ without shrinkage) are indicated by the black dots. The estimated coefficients for all $99$ noninformative covariates are visualized in a single boxplot for each state.}
\label{fig3}
\end{minipage}
\hfill
\begin{minipage}{0.4875\textwidth}
\includegraphics[width=\textwidth]{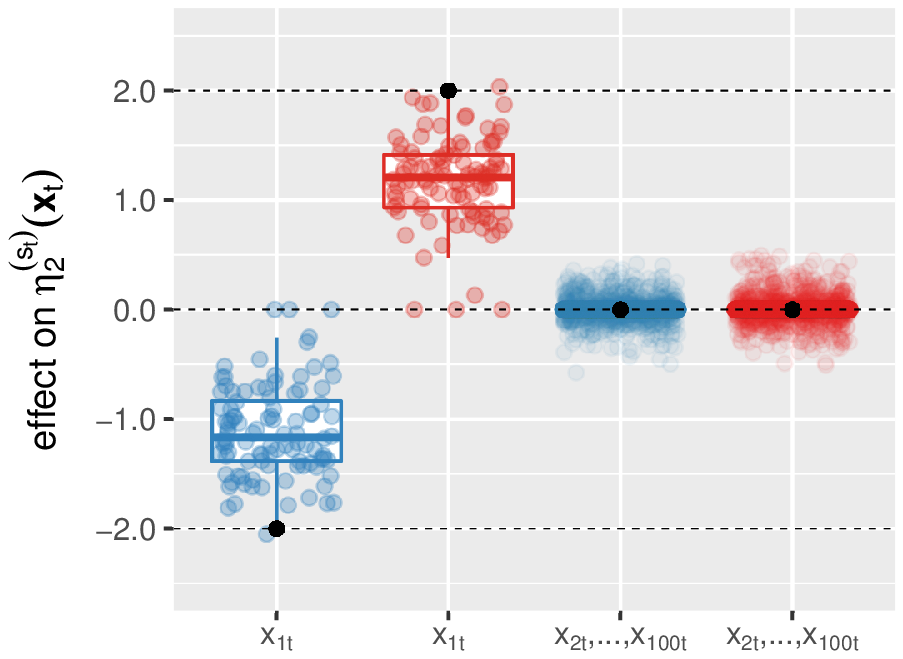}
\caption{Estimated state-dependent coefficients for $\theta_{2t}^{(s_t)}$.}
\vspace*{30mm}
\label{fig4}
\end{minipage}
\end{figure}

The sample means of the estimated off-diagonal t.p.m.\ entries, $\hat \gamma_{12}$ and $\hat \gamma_{21}$, were obtained as $0.047$ (standard deviation: $0.020$) and $0.047$ ($0.019$), respectively, which apparently is very close to the true values. The estimated state-dependent coefficients obtained in $100$ simulation runs are displayed in Figures \ref{fig3} and \ref{fig4}, respectively: For $\theta_{1t}^{(s_t)}$, the estimated coefficients are slightly shrunken towards zero, while for $\theta_{2t}^{(s_t)}$, the shrinkage effect is quite large.
The informative covariates were --- on average --- selected in $98.5$ \% of the cases ($100.0$ \% for $\theta_{1t}^{(s_t)}$ and $97.0$ \% for $\theta_{2t}^{(s_t)}$), while the noninformative ones were --- on average --- selected in $10.6$ \% of the cases ($13.4$ \% for $\theta_{1t}^{(s_t)}$ and $7.7$ \% for $\theta_{2t}^{(s_t)}$), which on the one hand indicates that the variable selection works quite well but on the other hand that there is a tendency towards too many covariates being included in the model (this apparently is a problem related to boosting in general rather than a specific one related to the MS-gamboostLSS algorithm, see e.g.\ the simulation experiments presented in \citealp{may12}). 

Using a 3.6 GHz Intel\textregistered{ }Core\texttrademark{ }i7 CPU, the average computation time was $1.4$ minutes for a (single) model (i.e.\ for a given number of boosting iterations), which is remarkably fast considering the fact that it involves variable selection among $100$ potential explanatory variables.

% nonlinear setting
\subsection{Nonlinear setting}
\label{sec4.2}

Encouraged by the performance in the linear setting, we next present a nonlinear setting using P-splines as base-learners, again simulating $500$ realizations from a $2$-state Markov chain with off-diagonal t.p.m.\ entries $\gamma_{ij} = 0.05$, $i,j=1,2$, $i\neq j$ and initial state probabilities $\delta_i=0.5$, $i,j=1,2$. We then draw $500$ observations from a normal distribution with state-dependent p.d.f.
\begin{equation*}
f_Y\left(y_t; \theta_{1t}^{(s_t)}, \theta_{2t}^{(s_t)}\right) = \frac{1}{\sqrt{2 \pi {\theta_{2t}^{(s_t)}}^2}} \exp \left( - \frac{\left(y_t - \theta_{1t}^{(s_t)}\right)^2}{2 {\theta_{2t}^{(s_t)}}^2} \right),
\end{equation*}
where
\begin{equation*}
\begin{split}
\theta_{1t}^{(1)} &= \eta_1^{(1)}(\mathbf{x}_t) = 2+2 \sin (\pi ( x_{1t}-0.5)) + \sum_{j=2}^{100} 0 x_{jt}\\
\theta_{1t}^{(2)} &= \eta_1^{(2)}(\mathbf{x}_t) = - 2 - \sin (\pi ( x_{1t}-0.5)) + \sum_{j=2}^{100} 0 x_{jt}\\
\log(\theta_{2t}^{(1)}) &= \eta_2^{(1)}(\mathbf{x}_t) = \sin (\pi (x_{1t}-0.5)) + \sum_{j=2}^{100} 0 x_{jt}  \\
\log(\theta_{2t}^{(2)}) &= \eta_2^{(2)}(\mathbf{x}_t) = - 2 \sin (\pi ( x_{1t}-0.5)) + \sum_{j=2}^{100} 0 x_{jt}
\end{split}
\end{equation*}
and $x_{jt} \sim \text{uniform}(-1,1)$, $j=1,\dots,100,t=1,\dots,500$. The stopping iterations were again chosen via $20$-fold cross validation over the grid $\boldsymbol{\Lambda}= \mathbf{n}_\text{stop}^{(1)} \times \mathbf{n}_\text{stop}^{(2)}$, $\mathbf{n}_\text{stop}^{(1)}=\mathbf{n}_\text{stop}^{(2)}= (25, 50, 100, 200)$, where the average chosen number of boosting iterations was $141.5$ (state 1) and $177$ (state 2).

\begin{figure}[t!]
\begin{minipage}{0.4875\textwidth}
\includegraphics[width=\textwidth]{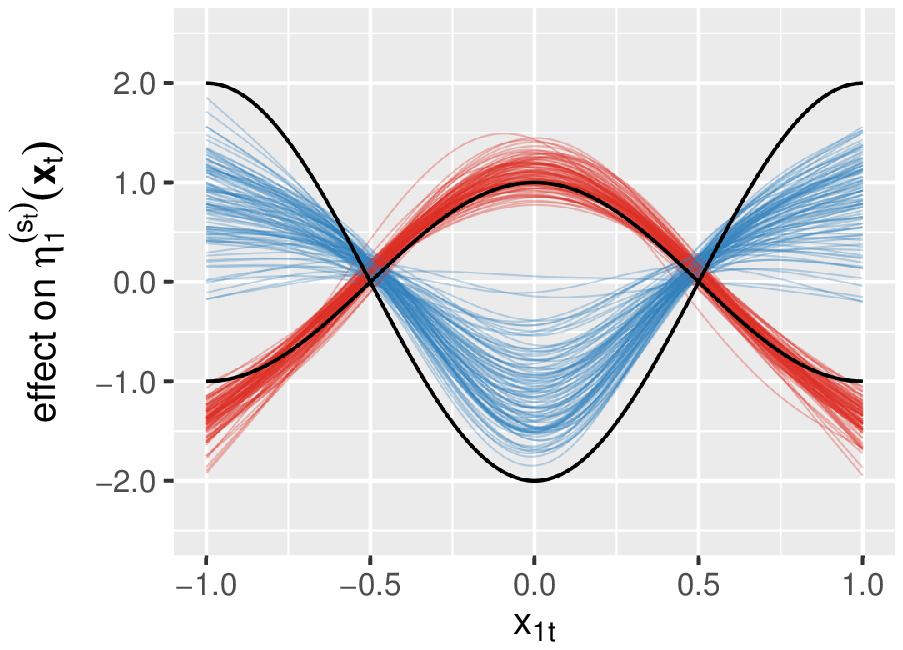}
\caption{Estimated state-dependent effects on $\theta_{1t}^{(s_t)}$ for state 1 (blue) and 2 (red) obtained in $100$ simulation runs. The true effects (i.e.\ without shrinkage) are indicated by the black lines. All effects have been centered around $0$.}
\label{fig5}
\end{minipage}
\hfill
\begin{minipage}{0.4875\textwidth}
\includegraphics[width=\textwidth]{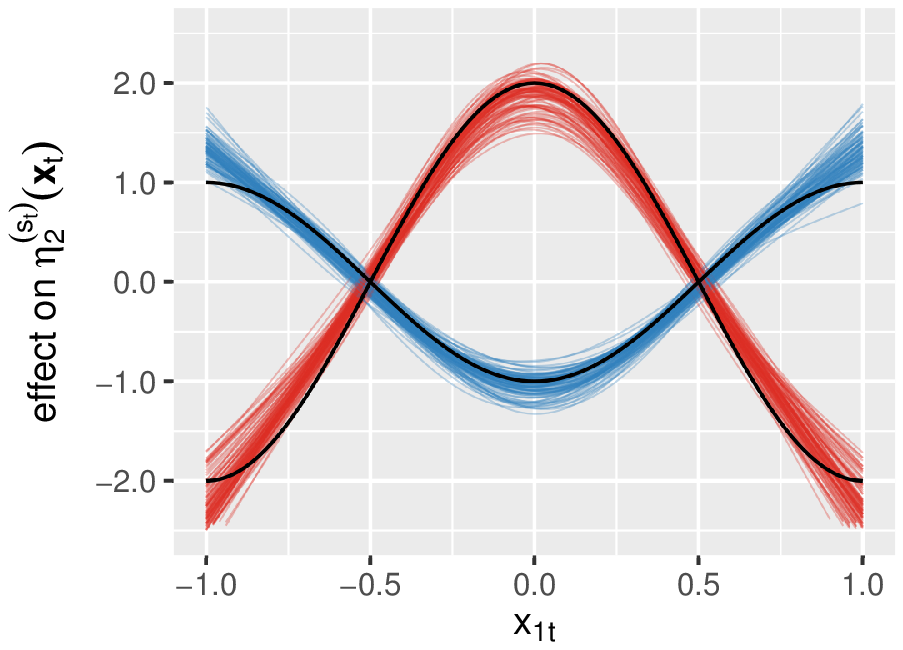}
\caption{Estimated state-dependent effects on $\theta_{2t}^{(s_t)}$.}
\vspace*{20.25mm}
\label{fig6}
\end{minipage}
\end{figure}

The sample means of the estimated off-diagonal t.p.m.\ entries, $\hat \gamma_{12}$ and $\hat \gamma_{21}$, were obtained as $0.050$ ($0.014$) and $0.051$ ($0.016$), respectively. The estimated state-dependent effects obtained in $100$ simulation runs are displayed in Figures \ref{fig5} and \ref{fig6}, respectively: As in Section \ref{sec4.1}, we observe a shrinkage effect (especially for the larger effects, i.e.\ the effects of $x_{1t}$ on $\eta_1^{(1)}(\mathbf{x}_t)$ and $\eta_2^{(2)}(\mathbf{x}_t)$); in addition, a smoothing effect can be observed (particularly for very small and large values of $x_{1t}$). The informative covariates were selected in all cases, while the noninformative ones were --- on average --- selected in $11.2$ \% of the cases ($7.4$ \% for $\theta_{1t}^{(s_t)}$ and $15.0$ \% for $\theta_{2t}^{(s_t)}$), which again indicates that the variable selection works quite well but apparently is not very conservative (particularly in the case of $\theta_{2t}^{(s_t)}$, where the shrinkage effect is considerably smaller than the one for $\theta_{1t}^{(s_t)}$, the average number of noninformative explanatory variables included in the model is fairly large).

For a given number of boosting iterations, model fitting took --- on average --- $7.8$ minutes per (single) model, which again is quite remarkable considering the fact that it does not only involve variable selection among $100$ potential covariates (as in the linear setting) but also results in smooth fits (without relying on a computer-intensive smoothing parameter selection).

%%% section 5 -- energy prices in spain %%%
\section{Energy prices in Spain}
\label{sec5}

To illustrate the suggested approach in a real-data setting, we model the conditional distribution of the daily average price of energy in Spain (in Cents per kWh), $Y_t$, over time. Our aim here is to present a simple case-study that provides some intuition and demonstrates the potential of MS-GAMLSS, which is why we focus on a relatively simple model involving only one explanatory variable, the daily oil price (in Euros per barrel), $x_{1t}$. The data, which are available in the R package MSwM \citep{san14}, cover $1761$ working days between February 1, 2002 and October 31, 2008. As in Section \ref{sec4.2}, we assume a normal distribution for the $Y_t$ and fitted two different $2$-state MS-GAMLSS with state-dependent predictors for the conditional mean, $\theta_{1t}^{(s_t)}$, and the conditional variance, $\theta_{2t}^{(s_t)}$, considering i) simple linear models (linear model), and ii) P-splines (nonlinear model) as base-learners. The stopping iterations were chosen via $20$-fold cross validation over the grid $\boldsymbol{\Lambda}= \mathbf{n}_\text{stop}^{(1)} \times \mathbf{n}_\text{stop}^{(2)}$, $\mathbf{n}_\text{stop}^{(1)}=\mathbf{n}_\text{stop}^{(2)}= (25, 50, 100, 200, 400, 800, 1600, 3200)$, which led to the optimal values $n_\text{stop}^{(1)}= 100$, $n_\text{stop}^{(2)} = 200$ (linear model) and $n_\text{stop}^{(1)}=1600$, $n_\text{stop}^{(2)} = 200$ (nonlinear model). For the chosen stopping iterations, the computation times were $0.4$ minutes (linear model) and $12.9$ minutes (nonlinear model).

\begin{figure}[t!]
\begin{minipage}{0.4875\textwidth}
\includegraphics[width=\textwidth]{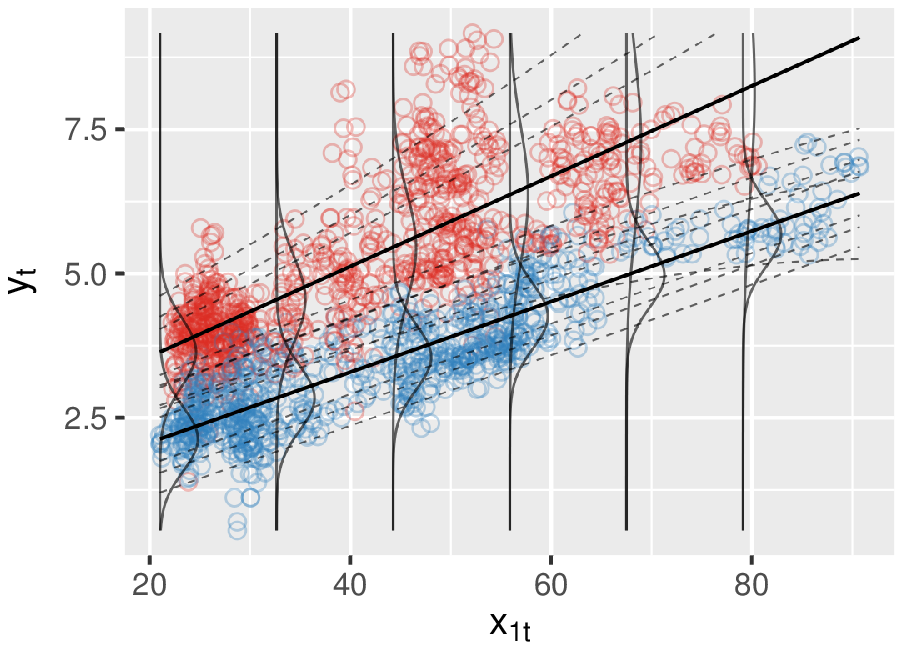}
\includegraphics[width=\textwidth]{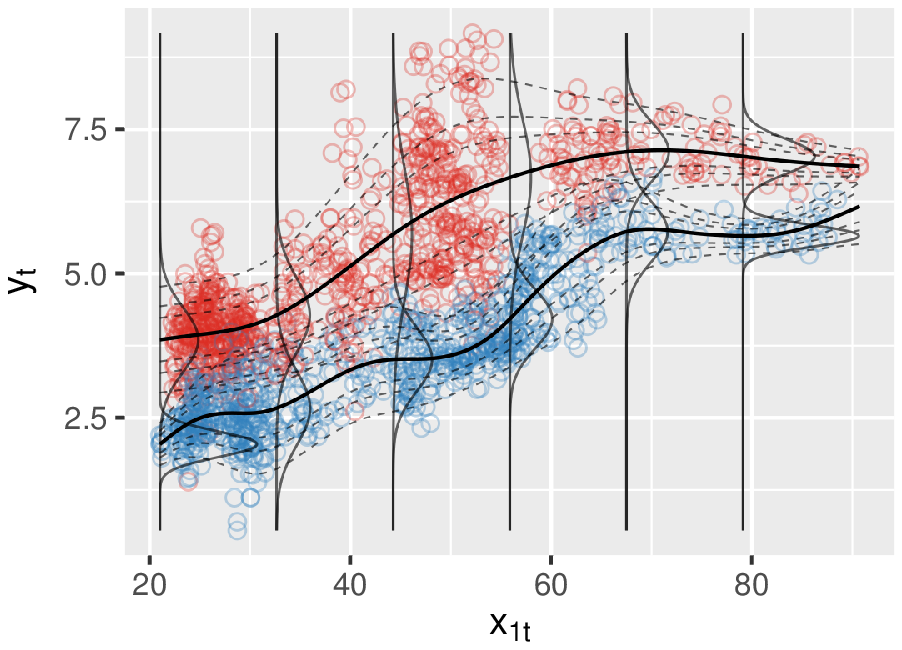}
\caption{Estimated state-dependent predictors for the mean (horizontal solid lines) for states 1 (blue) and 2 (red) and fitted state-dependent distributions for different values of $x_{1t}$ (vertical solid lines), which were computed based on the estimated state-dependent predictors for the variance. Dashed lines indicate the $0.05$, $0.15$, $0.25$, $0.75$, $0.85$ and $0.95$ quantiles of the fitted state-dependent distributions.}
\label{fig7}
\end{minipage}
\hfill
\begin{minipage}{0.4875\textwidth}
\includegraphics[width=\textwidth]{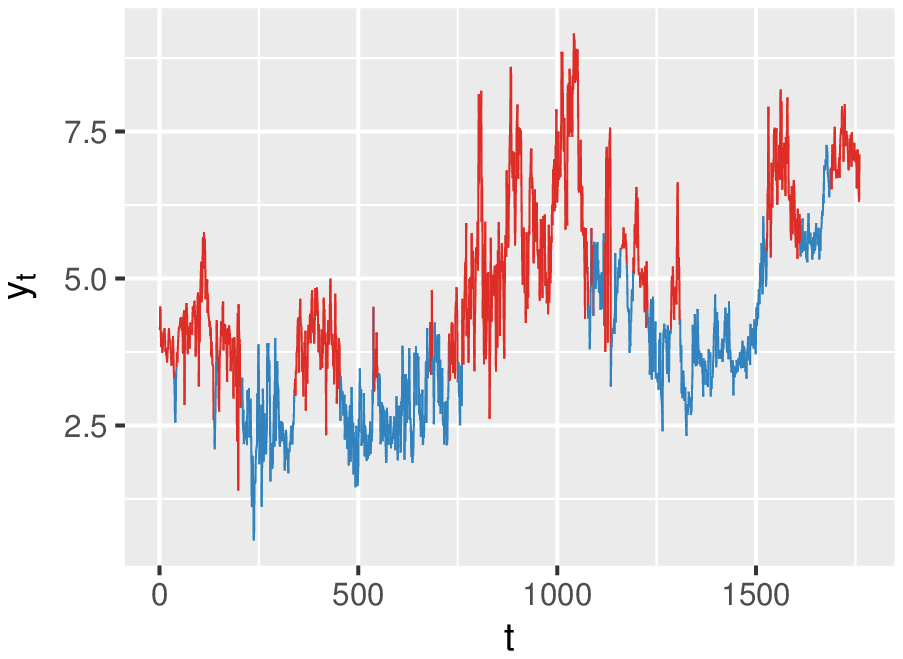}
\includegraphics[width=\textwidth]{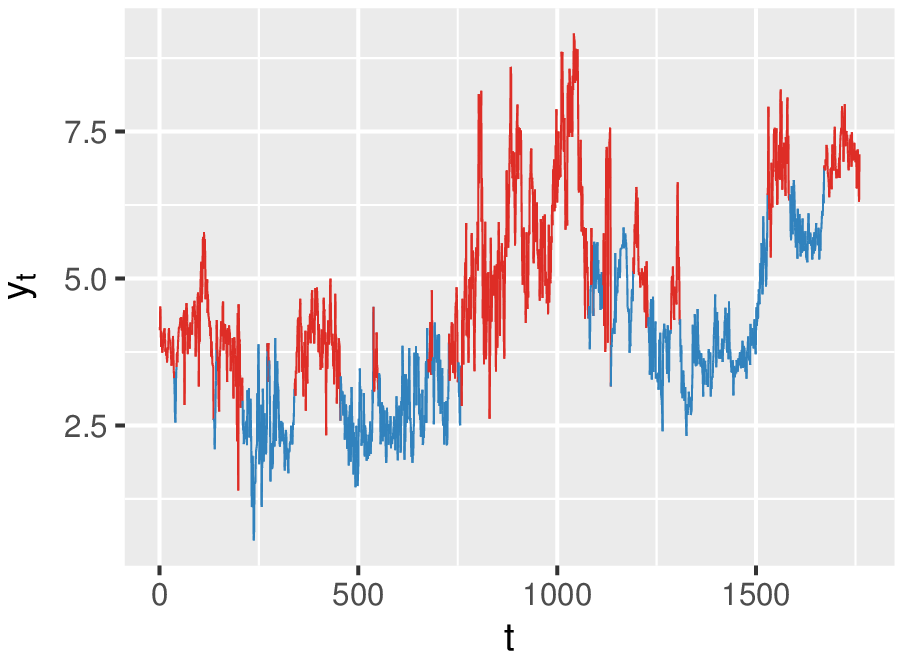}
\caption{Locally decoded time series of the daily energy prices.\vspace*{40.55mm}}
\label{fig8}
\end{minipage}
\end{figure}

The off-diagonal t.p.m.\ entries were estimated as $\hat \gamma_{12} = 0.017$, $\hat \gamma_{21} = 0.016$ (linear model) and $\hat \gamma_{12} = 0.020$, $\hat \gamma_{21} = 0.018$ (nonlinear model), which in both cases indicates high persistence within the states (according to the fitted models, the average dwell-times within a state were --- depending on the model and the state --- between $50$ and $62.5$ days). The estimated state-dependent distributions, as well as the locally decoded time series of the daily energy prices, are visualized in Figures \ref{fig7} and \ref{fig8}, respectively: According to both models, the oil price exhibits a (mostly) positive effect on the conditional mean, which essentially holds for both states. However, the linear model lacks the flexibility to capture the decreasing effect for $x_{1t} \geq 60$ that is revealed by the nonlinear model, which leads to a severe overestimation in that area. The effect on the conditional variance considerably differs across the states: In state $1$, the oil price has only a minor effect, whereas in state $2$, the conditional variance is strongly affected by the oil price. As in the case of the conditional mean, the effect on the conditional variance clearly has a nonlinear form (the volatility is relatively high for $40 \leq x_{1t} \leq 60$ and relatively low for $60 \leq x_{1t} \leq 40$), which is well-captured by the nonlinear model but not captured by the linear model. The consequence is a severe under- (over-) estimation for $40 \leq x_{1t} \leq 60$ ($40 \geq x_{1t} \geq 60$), as indicated by the quantile curves for the linear model depicted in Figure \ref{fig7}. From an economic point of view, state 1 may be linked to a calm market regime (which implies relatively low prices alongside a moderate volatility). State 2, in contrast, may correspond to a nervous market (which implies relatively high prices alongside a high volatility).

The results clearly demonstrate the potential of MS-GAMLSS: By accounting for the state-switching dynamics in the model formulation, they allow to draw a precise picture of the response distribution at any point in time, which may particularly be useful in applications where the focus lies on short-term forecasting. Furthermore, a precise picture of the entire response distribution (which certainly includes not only the mean, but also variance and potentially skewness and kurtosis parameters) is crucial when the focus is shifted from the expected value towards the quantiles, which for example is the case in risk measurement and portfolio optimization applications \citep{ace02}: Estimating the value-at-risk of a given investment, for instance, requires the prediction of certain quantiles of the corresponding loss distribution \citep{roc02}, which could potentially be addressed using MS-GAMLSS.

%%% section 6 -- discussion %%%
\section{Discussion}

We have introduced MS-GAMLSS as a novel class of flexible latent-state time series regression models which allows to model different parameters of the response distribution as potentially smooth functions of a given set of explanatory variables. Limitations of gradient boosting, particularly the fact that the design of the algorithm does not allow to compute standard errors for the effect estimates, also apply to the MS-gamboostLSS algorithm. While we have assumed a relatively simple state architecture, the underlying dependence structure could potentially be extended in various ways: i) higher-order Markov chains could be used to allow the states to depend not only on the previous state but on a sequence of multiple previously visited states \citep{zuc16}, ii) semi-Markov state processes could be used to specify arbitrary dwell-time distributions for the states \citep{lan11}, and iii) hierarchical state processes could be used to infer states at multiple temporal scales (\citealt{ada17}; \citealt{leo17}). Another potential feature of the latter approach is that multiple data streams collected at different time scales could be included in a joint, multivariate MS-GAMLSS, which may particularly be useful in economic applications, where data often tend to be collected on a daily, monthly or quarterly basis.

On a final note, we would like to raise awareness of the fact that the flexibility of MS-GAMLSS can be both a blessing and a curse: In some applications, MS-GAMLSS could potentially be overparameterized, and models as complex as MS-GAMLSS may not be appropriate even if they fit the data well (particularly in the case of short time series, overfitting may become a severe problem). It is therefore worth mentioning that MS-GAMLSS contain other, nested (i.e.\ less complex) HMM-type models, e.g.\ simple HMMs \citep{zuc16} or Markov-switching (generalized) linear and additive models (\citealp{lan17}; \citealp{lan18}). By specifying appropriate base-learners, the MS-gamboostLSS algorithm can be used to fit all these nested special cases: Using intercept-only terms (hence neglecting any covariate dependence), for instance, results in simple HMMs, while using simple linear models or P-splines for the conditional mean and intercept-only terms for the other parameters leads to Markov-switching (generalized) linear and additive models, respectively. Since none of the latter classes of models has been incorporated into the gradient boosting framework yet, the MS-gamboostLSS algorithm, which lies at the core of this work, may provide a promising method for model fitting and variable selection not only in MS-GAMLSS but also in a variety of other HMM-type models.

%%% references %%%
\renewcommand\refname{References}
\makeatletter
\renewcommand\@biblabel[1]{}

\end{spacing}

\end{document}